\documentclass[manuscript]{aastex}

\shorttitle{BLACK HOLE MASSES AND TIMESCALES IN BLAZARS}

\shortauthors{Liu \& Bai}

\begin{document}

\title{CONSTRAINTS ON BLACK HOLE MASSES WITH TIMESCALES OF VARIATIONS IN BLAZARS}

\author{H. T. Liu\altaffilmark{1,2} and J. M. Bai\altaffilmark{1,2}}

\altaffiltext{1}{Yunnan Observatories, Chinese Academy of
Sciences, Kunming, Yunnan 650011, China}

\altaffiltext{2}{Key Laboratory for the Structure and Evolution of
Celestial Objects, Chinese Academy of Sciences, Kunming, Yunnan
650011, China}

\email{htliu@ynao.ac.cn}

\begin{abstract}
In this paper, we investigated the issue of black hole masses and
minimum timescales of jet emission for blazars. We proposed a
sophisticated model that sets an upper limit to the central black
hole masses $M_{\bullet}$ with the minimum timescales $\Delta
t^{\rm{ob}}_{\rm{min}}$ of variations observed in blazars. The
value of $\Delta t^{\rm{ob}}_{\rm{min}}$ presents an upper limit
to the size of blob in jet. The blob is assumed to be generated in
the jet-production region in the vicinity of black hole, and then
the expanding blob travels outward along the jet. We applied the
model to 32 blazars, 29 of which were detected in gamma rays by
satellites, and these $\Delta t^{\rm{ob}}_{\rm{min}}$ are on the
order of hours with large variability amplitudes. In general,
these $M_{\bullet}$ estimated with this method are not
inconsistent with those masses reported in the literatures. This
model is natural to connect $M_{\bullet}$ with $\Delta
t^{\rm{ob}}_{\rm{min}}$ for blazars, and seems to be applicable to
constrain $M_{\bullet}$ in the central engines of blazars.

\end{abstract}


\section{INTRODUCTION}
Blazars are radio-loud active galactic nuclei (AGNs), including BL
Lacertae objects (BL Lac objects) and flat spectrum radio quasars
(FSRQs), characterized with some special observational features,
such as luminous nonthermal continuum emission from radio up to
GeV/TeV energies, rapid variability with large amplitudes, and
superluminal motion of their compact radio cores
\citep[e.g.][]{Ur95}. These unusual characteristics originate from
the Doppler boosted emission of a relativistic jet with a small
angle to the line of sight \citep{Bl78}. The intranight or
intraday variability (IDV) is an intrinsic phenomenon, and tightly
constrain the diameters of the emitting regions in blazars
\citep{WW95}. The emitting region in the relativistic jet is
usually simplified as a blob. The size $D$ of blob can be limited
by $D \la \delta \Delta t_{\rm{ob}}c/(1+z)$, where $\delta$ is the
Doppler factor of jet, $\Delta t_{\rm{ob}}$ is the observed
timescale of variations, $z$ is the redshift of source, and $c$ is
the speed of light. The IDV limit $D$ to be smaller than the size
of solar system by $\Delta t <$ 1 day in the source rest frame.
These timescales of the variations with large amplitudes in the
optical--gamma-ray bands might have some underlying connection
with the black hole masses of the central engines in blazars.

The relativistic jets can be generated from inner accretion disk
in the vicinity of black hole \citep[e.g.][]{Pe69,Bl77,Bl82,Me01}.
Observations show that dips in the X-ray emission, generated in
the central engines, are followed by ejections of bright
superluminal radio knots in the jets of AGNs and microquasars
\citep[e.g.][]{Ma02,Ar10,Ch09,Ch11}. The dips in the X-ray
emission are well correlated with the ejections of bright
superluminal knots in the radio jets of  3C 120 \citep{Ch09} and
3C 111 \citep{Ch11}. An instability in accretion flow may cause a
section of the inner disk break off, and the loss of this section
leads to a decrease in the soft X-ray flux, observed as a dip in
the X-ray emission. A fraction of the section is accreted into the
event horizon of the centra black hole. A considerable portion of
the section is ejected into the jet, observed as the appearance of
a superluminal bright knot. General relativistic
magnetohydrodynamic simulations showed production of relativistic
jets in the vicinity of black hole with jet-production region
around 7--8$r_{\rm{g}}$ for the Schwarzchild black hole and being
of the order of the radius of the ergosphere
$r_{\rm{e}}=2r_{\rm{g}}$ for the Kerr black hole, where
$r_{\rm{g}}$ is the gravitational radius of black hole
\citep{Me01}. For the Schwarzchild black hole, the inner radius of
the accretion disk is about equal to that of the marginally stable
orbit 6$r_{\rm{g}}$, which is comparable to the size of the
jet-production region. Thus the initial size of a blob emerged
from the jet-production region may be comparable to that of the
marginally stable orbit or that of the ergosphere. The blob will
expand as it travels outward along the jet. When the blob pass the
site of dissipation region in the jet, it will produce the
corresponding variations in the optical--gamma-ray regimes. The
minimum timescales of the variations are likely to be related with
the masses of the central black holes. Then the black hole masses
$M_{\bullet}$ can be constrained with the observed minimum
timescales $\Delta t_{\rm{min}}^{\rm{ob}}$ of variations in the
optical--gamma-ray regimes. In this paper, we attempt to construct
a model that constrains $M_{\bullet}$ with $\Delta
t_{\rm{min}}^{\rm{ob}}$ for blazars.

The structure of this paper is as follows. Section 2 presents
method. Section 3 presents applications. Section 4 is for
discussion and conclusions.

\section{METHOD}
Assuming the blob has a size of $D_{\rm{0}}$ in the jet-production
region (with a size comparable to $D_{\rm{0}}$), and a size of
$D_{\rm{R}}$ at the location $R_{\rm{jet}}$ in the jet from the
central engine (see Figure 1), we have an equation between
$D_{\rm{0}}$ and $D_{\rm{R}}$
\begin{equation}
D_{\rm{R}}=D_{\rm{0}}+2\overline{v}_{\rm{exp}}\frac{R_{\rm{jet}}}{\overline{v}_{\rm{jet}}}
=D_{\rm{0}}+2R_{\rm{jet}}\frac{\overline{v}_{\rm{exp}}}{\overline{v}_{\rm{jet}}},
\end{equation}
where $\overline{v}_{\rm{exp}}$ is the average expansion velocity
of blob in the jet between the central engine and the location
$R_{\rm{jet}}$, and $\overline{v}_{\rm{jet}}$ is the corresponding
average bulk velocity of the blob. The jet velocity
$\overline{v}_{\rm{jet}}$ is in a relativistic region. The
expansion velocity $\overline{v}_{\rm{exp}}$ is not in a
relativistic region. Then $\overline{v}_{\rm{exp}}\ll
\overline{v}_{\rm{jet}}$, and we have
\begin{equation}
D_{\rm{0}}\la D_{\rm{R}}.
\end{equation}
\begin{figure}[htp]
\begin{center}
\includegraphics[angle=-90,scale=0.4]{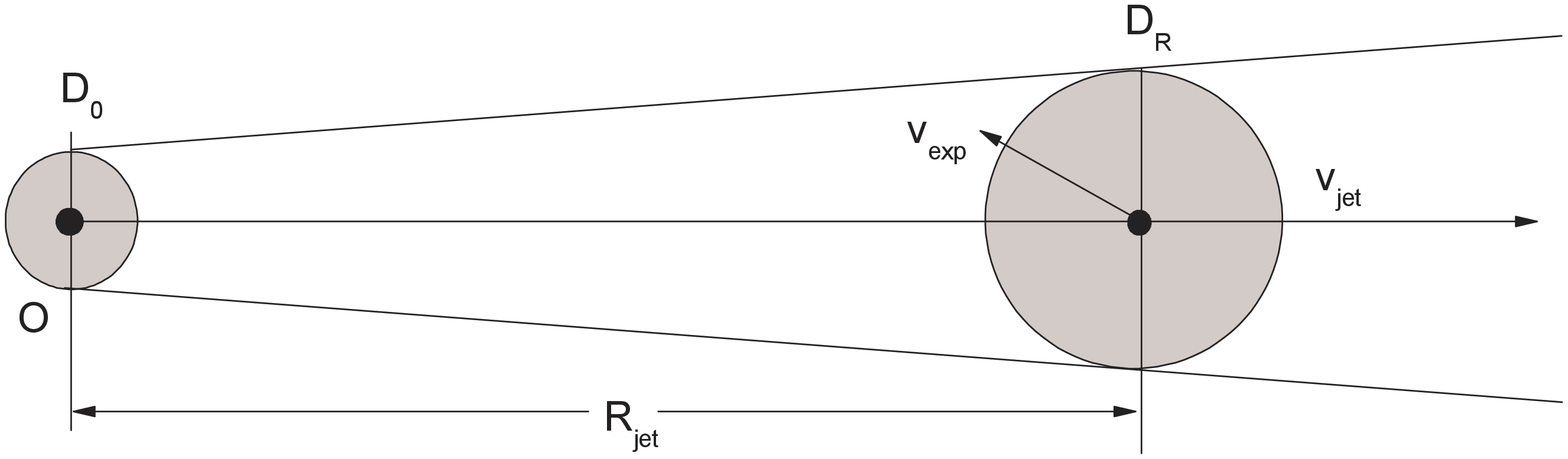}
\end{center}
 \caption{Sketch of axial cross section of geometry. $D_{\rm{0}}$ is the size of blob in the jet formation region around the central engine,
 and $D_{\rm{R}}$ is the size of blob at the position $R_{\rm{jet}}$ in the jet from the central engine.}
  \label{fig1}
\end{figure}
The size $D_{\rm{R}}$ of a blob at the location $R_{\rm{jet}}$ can
be constrained by the observed minimum timescale $\Delta
t_{\rm{min}}^{\rm{ob}}$ of the variations from the blob, and we
have
\begin{equation}
D_{\rm{R}}\leq \frac{\delta \Delta t_{\rm{min}}^{\rm{ob}}}{1+z}c,
\end{equation}
where $\delta$ is the Doppler factor, $z$ is the redshift of
source, and $c$ is the light speed. Combining equations (2) and
(3), we have
\begin{equation}
D_{\rm{0}}\la \frac{\delta \Delta t_{\rm{min}}^{\rm{ob}}}{1+z}c.
\end{equation}

The inner radius of accretion disk is usually taken to be around
the marginally stable orbit of disk surrounding the central black
hole. The radius of marginally stable orbit of disk is
\begin{equation}
r_{\rm{ms}}=r_{\rm{g}}\{3+Z_2 \mp [(3-Z_1)(3+Z_1+2Z_2 )]^{1/2}\},
\end{equation}
where $r_{\rm{g}}=GM_{\bullet}/c^2$ is the gravitational radius of
black hole with a mass of $M_{\bullet}$, $Z_1 \equiv
1+(1-j^2)^{1/3}[(1+j)^{1/3}+(1-j)^{1/3}]$, and $Z_2 \equiv
(3j^2+Z_1^2)^{1/2}$ \citep{Ba72}. Here, $j=J/J_{\rm{max}}$ is the
dimensionless spin parameter of black hole with the maximum
possible angular momentum $J_{\rm{max}}=GM_{\rm{\bullet}}^2/c$
with $G$ being the gravitational constant. In the case of the
prograde rotation, $r_{\rm{ms}}=6r_{\rm{g}}$ for $j=0$ and
$r_{\rm{ms}}=r_{\rm{g}}$ for $j=1$. General relativistic
magnetohydrodynamic simulations show that in the Schwarzchild case
the jet-production region has a size around 14--16$r_{\rm{g}}$
\citep{Me01}, comparable to the diameter of the marginally stable
orbit, $D_{\rm{ms}}=12 r_{\rm{g}}$. In the Kerr case, the
jet-production region must be of the order of the diameter of the
ergosphere $D_{\rm{e}}=2r_{\rm{e}}=4r_{\rm{g}}$, where
$r_{\rm{e}}$ is the equatorial boundary of the ergosphere
\citep{Me01}. Thus the jet-production region span about from
$D_{\rm{e}}$ to $D_{\rm{ms}}$ for $0\leq j\la 1$, and then we take
$D_{\rm{0}}=$4--12$r_{\rm{g}}$. From equation (4) and
$D_{\rm{0}}=$4--12$r_{\rm{g}}$, we have
\begin{mathletters}
\begin{eqnarray}
M_{\bullet}\la 5.086\times 10^4 \frac{\delta \Delta
t_{\rm{min}}^{\rm{ob}}}{1+z}M_{\odot} \/\ \/\ (D_{\rm{0}}=D_{\rm{e}}, \/\ j\sim 1),\\
M_{\bullet}\la 1.695 \times 10^4 \frac{\delta \Delta
t_{\rm{min}}^{\rm{ob}}}{1+z}M_{\odot} \/\ \/\
(D_{\rm{0}}=D_{\rm{ms}}, \/\ j=0),
\end{eqnarray}
\end{mathletters}
where $\Delta t_{\rm{min}}^{\rm{ob}}$ is in units of seconds.
Equations (6a) and (6b) can be unified as
\begin{equation}
M_{\bullet}\la 1.695{-}5.086 \times 10^4 \frac{\delta \Delta
t_{\rm{min}}^{\rm{ob}}}{1+z}M_{\odot}
\end{equation}

\section{APPLICATIONS}
The new method is applied to 32 blazars. These blazars have
variability timescales of the order of hours with large
variability amplitudes and a redshift range from 0.031 to 1.813.
These timescales were reported in the literatures for the
optical--X-ray--gamma-ray bands. The satellites detected GeV gamma
rays from 29 out of 32 blazars. The details of these blazars are
presented in Table 1. The minimum timescales of gamma rays are
taken to be the doubling times of fluxes. The optical minimum
timescales were reported in the literatures presented in column
(6) of Table 1. The doubling times of X-ray fluxes are taken as
the X-ray minimum timescales. These observed minimum timescales
are listed in column (4), and their corresponding references are
presented in column (6). The optical--$\gamma$-ray emission are
mostly the Doppler boosted emission of jets for gamma-ray blazars
\citep{Gh98}. A value of $\delta \sim$ 10 was adopted for GeV
gamma-ray blazars \citep{Gh10}. We will take $\delta =10$ to
estimate $M_{\bullet}$ with formula (7).

The Schwarzchild and Kerr black holes are considered in estimates
of $M_{\bullet}$ from formula (7). The estimated black hole masses
are denoted by $M^{\rm{Ker}}_{\rm{var}}$ and
$M^{\rm{Sch}}_{\rm{var}}$ in columns (7) and (8) of Table 1,
respectively. In the Schwarzchild case, $M^{\rm{Sch}}_{\rm{var}}$
spans from $10^{8.13}$ to $10^{9.68}$ $M_{\rm{\odot}}$. In the
Kerr case, $M^{\rm{Ker}}_{\rm{var}}$ spans from $10^{8.61}$ to
$10^{10.16}$ $M_{\rm{\odot}}$. We compared these estimated
$M^{\rm{Sch}}_{\rm{var}}$ to those masses $M_{\rm{BH}}$ obtained
with other methods reported in the literatures. In Figure 2 of
$M_{\rm{BH}}$ versus $M^{\rm{Sch}}_{\rm{var}}$, there are five
blazars below the line of $M_{\rm{var}}=M_{\rm{BH}}$. Because
$M^{\rm{Sch}}_{\rm{var}}$ is only an upper limit of $M_{\bullet}$,
the area below this line is not allowed for these blazars. When
the central black holes are the Kerr ones, these upper limits of
masses are increased by a factor 3, and four out of five blazars
are moved into the area above the line of
$M_{\rm{var}}=M_{\rm{BH}}$. Only 4C +38.41 is just below the line
of $M_{\rm{var}}=M_{\rm{BH}}$. The populations of $M_{\bullet}$ in
Figure 2 show that these $M_{\rm{var}}$ estimated from formula (7)
are reasonable upper limits on $M_{\bullet}$ for blazars.
\begin{figure}
\begin{center}
\includegraphics[angle=-90,scale=0.4]{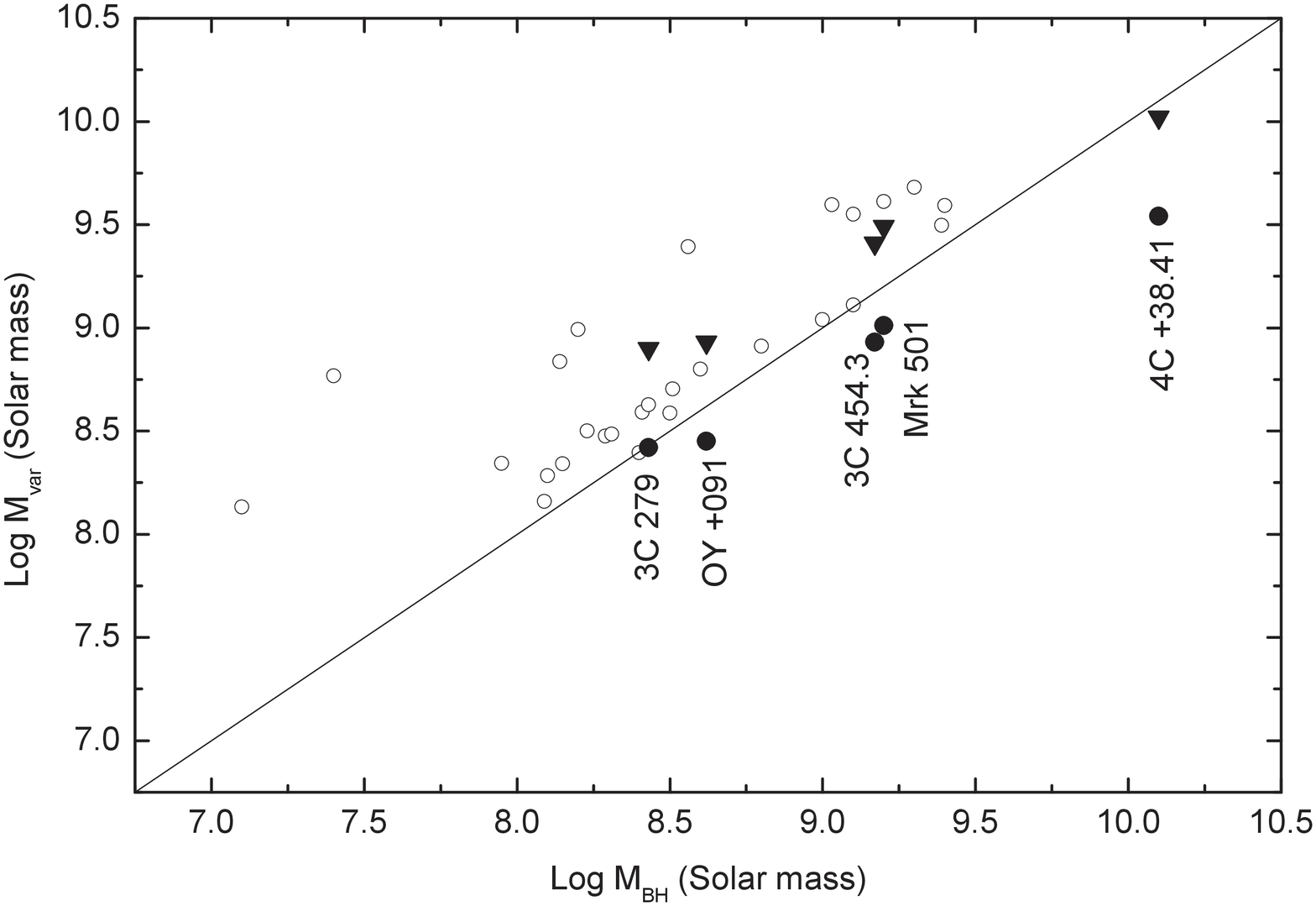}
\end{center}
 \caption{Black hole mass $M_{\rm{var}}$ versus $M_{\rm{BH}}$ estimated in other methods in the literatures.
 The solid line is $M_{\rm{var}}=M_{\rm{BH}}$. The triangles are $M_{\rm{var}}$ in the Schwarzchild case, and the
solid circles are $M_{\rm{var}}$ in the Kerr case.}
  \label{fig2}
\end{figure}

\section{DISCUSSION AND CONCLUSIONS}
\citet{Mo86} thought X-ray emission to be produced very close to
the inner engine in BL Lacertae object PKS 2155-304. \citet{Ah07}
limited the Doppler factor by the black hole mass and the
variability timescale of very high energy gamma-ray flare of PKS
2155-304. \citet{Mi89} reported the rapid variations on timescale
as short as 1.5 hours for BL Lacertae in the optical flux, and the
minimum timescale for the variations was used to place constraints
on the size of the emitting region. They assumed that these
variations were produced in the vicinity of a supermassive black
hole, and then determined a black hole mass with the minimum
timescale from a formula
\begin{equation}
M_{\bullet}= \frac{c^3 \Delta t_{\rm{min}}^{\rm{ob}}}{6G(1+z)},
\end{equation}
for the Schwarzchild black hole. At that time, they thought that
relativistic beaming need not be invoked to account for the
luminosity of this object. For the Kerr black hole, \citet{Xi02c}
deduced a formula from \citet{Ab82}
\begin{equation}
M_{\bullet}\la 1.62 \times 10^4 \frac{\Delta
t_{\rm{min}}^{\rm{ob}}}{1+z}M_{\odot},
\end{equation}
which gives an upper limit on $M_{\bullet}$. These above two
equations are based on assumption of accretion disk surrounding a
supermassive black hole, and the optical flux variations are from
the accretion disk. Obviously, these two formulae are applicable
to estimate $M_{\bullet}$ for non-blazar-like AGNs or some AGNs
with weaker blazar emission component in fluxes relative to
accretion disk emission component. Considering the relativistic
beaming effect, \citet{Xi02c} deduced a new formula from formula
(9)
\begin{equation} M_{\bullet}\la 1.62 \times 10^4 \frac{\delta
\Delta t_{\rm{min}}^{\rm{ob}}}{1+z}M_{\odot}.
\end{equation}
Formula (10) was applied to blazars, especially BL Lac objects
\citep[see][]{Xi02c,Xi05b}. In this paper, we proposed a
sophisticated model to constrain the black hole masses using the
rapid variations with large amplitudes for blazars. The model is
suitable to constrain $M_{\bullet}$ in blazars using the minimum
timescales of variations of the beamed emission from the
relativistic jet. Formula (7) is the same as formula (10) in
except of the coefficients in the right of formulae, but these two
corresponding models are essentially different in their origins.

Spectral energy distributions of blazars consist of two broad
peaks \citep[e.g.][]{Gh98}. The first peak is produced by the
synchrotron radiation processes of relativistic electrons in a
relativistic jet. The second one is generally believed to come
from the inverse Compton scattering processes of the same
population of electrons. In general, the optical--gamma-ray
emission are believed and/or assumed to be produced in the same
region, simplified as a blob in simulating the spectral energy
distributions of blazars. The coincidence of a gamma-ray flare
with a dramatic change of optical polarization angle provides
evidence for co-spatiality of optical and gamma-ray emission
regions in 3C 279 \citep{Ab10}. Thus the timescales of the
optical--X-ray--gamma-ray variations are used in formula (7). For
this model, the blob size is assumed to increase linearly as it
travels outward along the jet (see equation (1)). This linear
growth assumption is only an approximation to the actual growth.
The assumption will not change formula (2), and then don't alter
formula (7).

The errors associated with this method are quite high. Besides the
poorly known details of the model itself, related to our ignorance
of the exact distance from the black hole where the emission is
produced, a large uncertainty is related to the value of the
adopted Doppler factor. A value of $\delta = 10$ is assumed, which
may be in excess or short of the real value by at least a factor
3, resulting in a total uncertainty of at least an order of
magnitude. In fact, it is possible for 32 blazars listed in Table
1. The Doppler factor $\delta$ can be estimated from the Lorentz
factor and the viewing angle of jet adopted to model spectral
energy distributions of bright \textit{Fermi} blazars in
\citet{Gh10}. There are 21 \textit{Fermi} blazars with new
estimated $\delta$, and these new $\delta$ are from 13 to 28 with
an average of 17. So, the upper limits of black hole masses are
increased by a factor 1.3--2.8 when adopting these new $\delta$
for the 21 blazars. Thus the 29 \textit{Fermi} blazars out of 32
blazars we employed will have a similar case for the upper limits
of black hole masses. The adopted Doppler factor of $\delta =10$
may result in a large uncertainty with a factor 1.3--2.8 in the
upper limits of $M_{\bullet}$. Another larger uncertainty with a
factor 3.0 arises from the ignorance of spins of the central black
holes in blazars (see formulae (6) and (7)). Thus the two large
uncertainties will lead to errors of 3.9--8.4 in the upper limits
of $M_{\bullet}$ estimated with this method for blazars.

In this paper, we proposed a sophisticated model to constrain the
central black hole masses $M_{\bullet}$ with the observed minimum
timescales $\Delta t^{\rm{ob}}_{\rm{min}}$ of variations in
blazars. The size of a blob in the relativistic jet can be
constrained with $\Delta t^{\rm{ob}}_{\rm{min}}$. The blob is
assumed to be ejected from the jet-production region in the
vicinity of black hole, and then to expand linearly as it travels
outward along the jet. The model is applied to 32 blazars, out of
which 29 blazars were detected in the GeV gamma-ray regime with
the satellites. Their observed minimum timescales are on the order
of hours with large variability amplitudes in the
optical--X-ray--gamma-ray bands. In general, these $M_{\bullet}$
estimated from $\Delta t^{\rm{ob}}_{\rm{min}}$ by formula (7) are
not inconsistent with those masses reported in the literatures.
This indicates that this model is applicable to constrain
$M_{\bullet}$ in the central engines of blazars. This model is
more natural to connect $M_{\bullet}$ with $\Delta
t^{\rm{ob}}_{\rm{min}}$ for blazars. Due to the ignorance of the
black hole spins and the real Doppler factors, the uncertainty of
the upper limits of masses will be about 3.9--8.4 in this method.
The upper limits of masses will have uncertainty of 3.0 because
the black hole spins are unknown for blazars, though we have the
real Doppler factors rather than the adopted value of $\delta =
10$ in this paper.

\acknowledgements We are grateful to the anonymous referee for
constructive comments leading to significant improvement of this
paper. HTL thanks the National Natural Science Foundation of China
(NSFC; Grant 11273052) for financial support. JMB acknowledges the
support of the NSFC (Grant 11133006). HTL thanks the financial
support of the Youth Innovation Promotion Association, CAS and the
project of the Training Programme for the Talents of West Light
Foundation, CAS.

\clearpage

\begin{deluxetable}{rrrrrrrrrrr}

\tablecolumns{11}

\tabletypesize{\scriptsize}


\tablecaption{Sample of blazars \label{tbl-1}}

\tablewidth{0pt}

\tablehead{ \colhead{Blazar name} & \colhead{$z$} &
\colhead{Satelli.\tablenotemark{a}} & \colhead{$\Delta
t_{\rm{min}}$\tablenotemark{b}} & \colhead{Band} & \colhead{Ref.}
& \colhead{$M_{\rm{var}}^{\rm{Ker}}$\tablenotemark{c}} &
\colhead{$M_{\rm{var}}^{\rm{Sch}}$\tablenotemark{c}} &
\colhead{$M_{\rm{BH}}$\tablenotemark{c}} & \colhead{Ref.} \\

\colhead{(1)}&\colhead{(2)}&\colhead{(3)}&\colhead{(4)}&\colhead{(5)}&\colhead{(6)}&
\colhead{(7)}&\colhead{(8)}&\colhead{(9)}&\colhead{(10)} }

\startdata

3C 66A&0.444&EG,LAT&3.73&V&19&9.28&8.80&8.60&24\\
AO 0235+164&0.940&EG,LAT&4.74&R&9&10.16&9.68&9.30&24\\
RBS 0413&0.190&LAT&3.19&V&18&8.82&8.34&7.95&26\\
PKS 0420-01&0.916&EG,LAT&4.65&O&1&10.07&9.60&9.03&26\\
PKS 0537-441&0.894&EG,LAT&4.66&O&1&10.09&9.61&9.20&24\\
PKS 0548-322&0.069&None&3.14&R&21&8.82&8.34&8.15&22\\
S5 0716+71&0.300&EG,LAT&3.17&V&7&8.76&8.28&8.10&24\\
PKS 0735+17&0.424&EG,LAT&3.32&B&19&8.87&8.40&8.40&24\\
OJ 248 &0.941&EG,LAT&3.65&R&20&9.07&8.59&8.41&27\\
OJ 287&0.306&EG,LAT&3.59&R&19&9.18&8.70&8.51&23\\
MrK 421&0.031&EG,LAT&3.26&TeV&4&8.95&8.48&8.29&22\\
4C +29.45&0.725&EG,LAT&4.40&O&17&9.87&9.39&8.56&27\\
W Com&0.102&LAT&3.58&V&16&9.24&8.77&7.40&24\\
4C+21.35&0.432&EG,LAT&3.92&GeV&3&9.47&8.99&8.20&24\\

3C 273&0.158&EG,LAT&4.33&V&11&9.97&9.50&9.39&25\\
3C 279&0.536&EG,LAT&3.38&V&19&8.90&8.42&8.43&26\\
PKS 1406-076&1.494&EG,LAT&4.76&R&14&10.07&9.59&9.40&24\\
PKS 1510-089&0.360&EG,LAT&3.39&R&20&8.96&8.49&8.31&27\\
AP Lib &0.049&LAT&2.95&O&1&8.64&8.16&8.09&22\\
PKS 1622-29&0.815&EG,LAT&4.14&GeV&8&9.59&9.11&9.10&24\\
4C +38.41&1.813&EG,LAT&4.76&GeV&2&10.02&9.54&10.10&24\\
3C 345&0.593&None&3.60&I&19&9.10&8.63&8.43&27\\
MrK 501&0.034&LAT&3.80&R&19&9.49&9.01&9.20&22\\
3C 371 &0.051&LAT&3.38&X&12&9.06&8.59&8.50&22\\
PKS 2005-489&0.071&EG,LAT&3.84&X&5&9.52&9.04&9.00&26\\
OX 169&0.213&LAT&3.69&O&1&9.31&8.84&8.14&27\\
PKS 2155-304&0.116&EG,LAT&2.95&O&10&8.61&8.13&7.10&24\\
BL Lacertae&0.069&EG,LAT&3.30&B&15&8.98&8.50&8.23&26\\
CTA 102 &1.037&EG,LAT&4.63&O&1&10.03&9.55&9.10&24\\
3C 454.3&0.859&EG,LAT&3.97&O&13&9.41&8.93&9.17&26\\
OY +091 &0.190&None&3.30&B&15&8.93&8.45&8.62&26\\
1ES 2344+514&0.044&LAT&3.70&X&6&9.39&8.91&8.80&26\\

\enddata

\tablecomments{Column 1: Blazar names; Column 2: redshifts of
objects; Column 3: The $\gamma$-ray satellites detected $\gamma$
rays from objects; Column 4: the minimum timescales of variations
of objects; Column 5: the bands where the minimum timescales
measured, B: the {\it B} band; V: the {\it V} band; R: the {\it R}
band; I: the {\it I} band; O: the optical band; X: the X-ray band;
Column 6: the references for columns 4 and 5; Column 7: the Kerr
black hole masses estimated from equation (6a); Column 8: the
Schwarzchild black hole masses estimated from equation (6b);
Column 9: the black hole masses estimated with other methods in
the literatures; Column 10: the references for column 9. \\
\textbf{References}: (1) \citealt{Ba83}; (2) \citealt{Fa99}; (3)
\citealt{Fo11}; (4) \citealt{Ga96}; (5) \citealt{Gi90}; (6)
\citealt{Gi00}; (7) \citealt{Gu09}; (8) \citealt{Ma97}; (9)
\citealt{Mo85}; (10) \citealt{Pa97}; (11) \citealt{Ra11}; (12)
\citealt{St86}; (13) \citealt{Vi97}; (14) \citealt{Wa95}; (15)
\citealt{Xi90}; (16) \citealt{Xi91a}; (17) \citealt{Xi91b}; (18)
\citealt{Xi92}; (19) \citealt{Xi99}; (20) \citealt{Xi01}; (21)
\citealt{Xi02a}; (22) \citealt{Ba03}; (23) \citealt{Li02}; (24)
\citealt{Li03}; (25) \citealt{Pa05}; (26) \citealt{Wo02}; (27)
\citealt{Xi05a}.}

\tablenotetext{a}{Detected with equipments aboard $\gamma$-ray
satellites. EG: the EGRET experiment aboard the {\it Compton Gamma
Ray Observatory}, and LAT: the Large Area Telescope aboard the
{\it Fermi} satellite.}

\tablenotetext{b}{The logarithms of timescales are in units of
seconds. }

\tablenotetext{c}{The logarithms of black hole masses are in units
of Solar mass, $M_{\rm{\odot}}$.}

\end{deluxetable}

\clearpage

\end{document}